\documentclass[aps,pra,10pt]{revtex4-1} 
\usepackage{amsmath}
\usepackage{amsfonts}
\usepackage{textcomp}
\usepackage{graphicx}

\begin{document}

\title{Spatial nonlinearity in anisotropic metamaterial plasmonic slot waveguides}
\author{Mahmoud M. R. Elsawy}
\author{Gilles Renversez}
\affiliation{Aix--Marseille Univ, CNRS, Ecole Centrale Marseille, Institut Fresnel, 13013 Marseille, France}

\email[]{gilles.renversez@univ-amu.fr}

\date{\today}



\begin{abstract}
We study the main nonlinear solutions of plasmonic slot waveguides made from an anisotropic metamaterial core  with a positive Kerr-type  nonlinearity  surrounded by two semi-infinite metal regions.  First, we demonstrate that for a highly anisotropic diagonal elliptical core, the bifurcation threshold  of the asymmetric mode is reduced from GW/m threshold for the isotropic case to 50 MW/m one indicating a strong enhancement of the spatial nonlinear effects, and that the slope of the dispersion curve of the  asymmetric mode stays positive, at least near the bifurcation, suggesting a stable mode. Second, we show that for the hyperbolic case there is no physically meaningful asymmetric mode, and that the sign of the effective nonlinearity can become negative.
\end{abstract}

\maketitle
\thispagestyle{plain} 

Nonlinear plasmonics is now a thriving research field~\cite{Kauranen12}. Among it, its integrated branch where surface plasmon polariton waves propagates at least partially in nonlinear media is seen as promising in high-speed small footprint signal processing~\cite{diaz16kerr-effect-hybrid-plasmonic-waveguides}.
As a building block for nonlinear plasmonic circuitry, the nonlinear plasmonic slot waveguide (NPSW) is of crucial importance even in its simplest version~\cite{Feigenbaum07,Davoyan08}.  Since all the key features can be studied  and understood in detail, this structure allows us future generalizations from more complex linear structures like the hybrid plasmonic waveguide~\cite{oulton08hybrid-plasmonic-waveguide}.

The strong field confinement achieved by these plasmonic waveguides ensure a reinforcement of the nonlinear effects which can be boosted further using epsilon-near-zero (ENZ) materials as it was already shown~\cite{ciattoni_extreme_2010,ciattoni_enhanced_2016}. 
It is worth mentioning that metal nonlinearities have already been investigated including at least in one study  where the wavelength range of enhanced nonlinearity has been controlled using metamaterials~\cite{neira_eliminating_2015}. Here, we focus on structures where the nonlinearity is provided by dielectric materials like  hydrogenated amorphous silicon (a-Si:H)~\cite{Lacava13} due to its high intrinsic third order nonlinearity around the telecommunication wavelength and to its manufacturing capabilities. 

\begin{figure}[htbp]
\centerline{\includegraphics[width = 0.9\columnwidth,angle=0,clip=true,trim= 0 0 0 0]{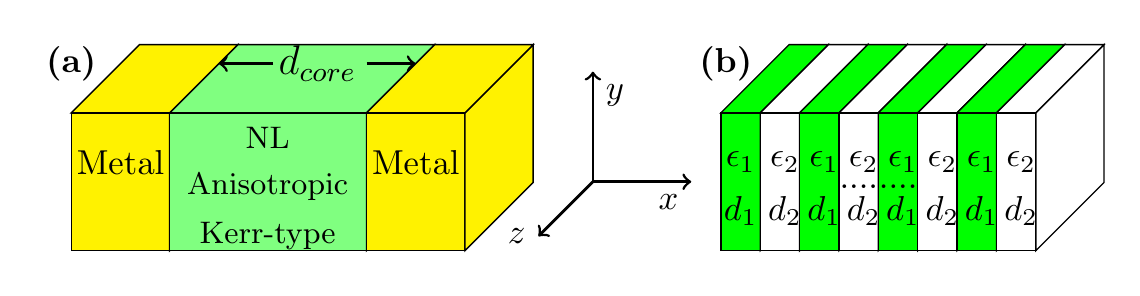}}
\caption{(a) Symmetric NPSW geometry with its metamaterial nonlinear core and the two semi-infinite metal regions. (b) Metamaterial nonlinear core obtained from a stack of two types of layers with permittivities and thicknesses $\varepsilon_1$ and $d_1$,  and  $\varepsilon_2$ and $d_2$, respectively. Only material 1 is nonlinear.}
\label{fig:geom-scheme-NPSW-metamat}
\end{figure}
In~\cite{ciattoni_extreme_2010}, nonlinear guided waves were investigated in anisotropic structures with an isotropic effective  dilectric response for transverse magnetic (TM) waves while here we consider a metamaterial core with an anisotropic effective dielectric response for TM waves. 
Other related works~\cite{neira_eliminating_2015,ciattoni_enhanced_2016} did not focus on  the nonlinear waveguide problem  or did not consider plasmonic structures. This task will be realized in the present study. 
 Furthermore, in indium tin oxide layer,  a record change of 0.72 in the refractive index increase induced by a third order nonlinearity has recently been reported~\cite{Alam795}.
  As concluded by the authors, this result challenges the usual hypothesis that the nonlinear term can be treated as a perturbation. One approach to tackle this problem, for the case of nonlinear stationary waves, is to take into account the spatial profile of the fields directly from Maxwell's equations as for example in~\cite{snyder94,Drouart08}. 
To start by this case is justified beacuse in waveguide studies whatever they are linear or nonlinear, it is well established that the modal approach is the first key step~\cite{Snyder:91} requiring to investigate the self-coherent stationary states of Maxwell's equations.
Here, extending methods we developed to study stationary states in isotropic NPSWs~\cite{Walasik15a,Walasik15b} to the anisotropic case, we describe the main properties obtained when a  nonlinear metamaterial  is used as core medium.
The examples of metamaterial nonlinear cores used in the following  are built using effective medium theory from well-known materials and realistic parameters.
The main nonlinear solutions in both elliptical and hyperbolic cases are investigated. 
In the first case, we demonstrate numerically and theoretically that for a highly anisotropic case, the effective nonlinearity~\cite{Walasik15a} can be enhanced nearly up to five orders of magnitude allowing a decrease of nearly three orders of magnitude of the bifurcation threshold of the asymmetric mode existing in  the symmetric structure~\cite{Davoyan08,Walasik15b}.
Next, we show that, in the hyperbolic case, changes appear in the field profiles compared to the simple isotropic NPSWs. We also demonstrate that due to the peculiar anisotropy, an effective  defocusing effect can be obtained from the initial positive Kerr nonlinearity. 

Figure~\ref{fig:geom-scheme-NPSW-metamat} shows a scheme of the nonlinear waveguide we investigate. 
 Compared to already studied NPSW with an isotropic nonlinear dielectric core~\cite{Feigenbaum07,Davoyan08,Walasik14a}, the new structure contains a metamaterial nonlinear core. We will study only symmetric structures  even if asymmetric isotropic NPSWs have already been considered~\cite{Walasik15b}.  
We consider monochromatic TM waves propagating along the $z$ direction (all field components evolve proportionally to $\exp[i( k_0 n_{eff} z - \omega t)]$) in a one-dimensional NPSW depicted in Fig.~\ref{fig:geom-scheme-NPSW-metamat}. Here $k_0 = \omega/c$, where $c$ denotes the speed of light in vacuum, $n_{eff}$ denotes the effective mode index and $\omega$ is the light angular frequency. The electric field comporents are $(E_{x},0,i E_z)$ and the magnetic field one is $(0,H_y,0)$.
In all the waveguide, the magnetic permeability is equal to $\mu_0$, the one of vacuum.

The nonlinear Kerr-type metamaterial core of thickness $d_{core}$ is anisotropic (see Fig.~~\ref{fig:geom-scheme-NPSW-metamat}). Its full effective permittivity tensor $\bar{\bar{\varepsilon}}_{eff}$ has only three non-null diagonal terms. Its  linear diagonal elements are $\varepsilon_{jj}$ $ \forall j \in \{x,y,z\} $.
We derive these terms from simple effective medium theory (EMT) applied to a stack of two isotropic material layers.
 $d_1$ and $d_2$ are the layer thicknesses of isotropic material 1 (nonlinear focusing Kerr-type) and material 2 (linear), respectively. Their respective linear permittivities are $\varepsilon_{1}$ and $\varepsilon_{2}$.
The EMT is typically valid when the light wavelength $\lambda$ is much larger than  $d_1$ and $d_2$. 
Depending on the chosen orientation of the compound layers relative to the Cartesian coordinate axes, different  anisotropic permittivity tensors can be build for the core. Due to the required $z$-invariance, only two types where the $z$-axis belongs to the layers have to be considered. For the first one where the layers are parallel to the $x$-axis, one has, for the linear diagonal terms of $\bar{\bar{\varepsilon}}_{eff}$:  $[ \varepsilon_{xx}  =  \varepsilon_{//} \;\,  \varepsilon_{yy}  =  \varepsilon_{\perp}   \;\,   \varepsilon_{zz}  =  \varepsilon_{//}] $ with $\displaystyle \varepsilon_{//} = \Re e ( r\varepsilon_{2}+(1-r) \varepsilon_{1})$, $\displaystyle  \varepsilon_{\perp}  = \Re e ( (\varepsilon_{1}\varepsilon_{2})/( r\varepsilon_{1}+(1-r)\varepsilon_{2}) )$, and $r= {d_{2}}/({d_{1}+d_{2}})$. For the second case  where the layers are parallel to the $y$-axis (see Fig.~\ref{fig:geom-scheme-NPSW-metamat} (b)), one gets:   $[ \varepsilon_{\perp}  \;\,  \varepsilon_{//}   \;\, \varepsilon_{//}] $. We will focus only on this case.

\begin{figure}[htbp]
\centerline{\includegraphics[width = 0.95\columnwidth,angle=0,clip=true,trim= 0 0 0 0]{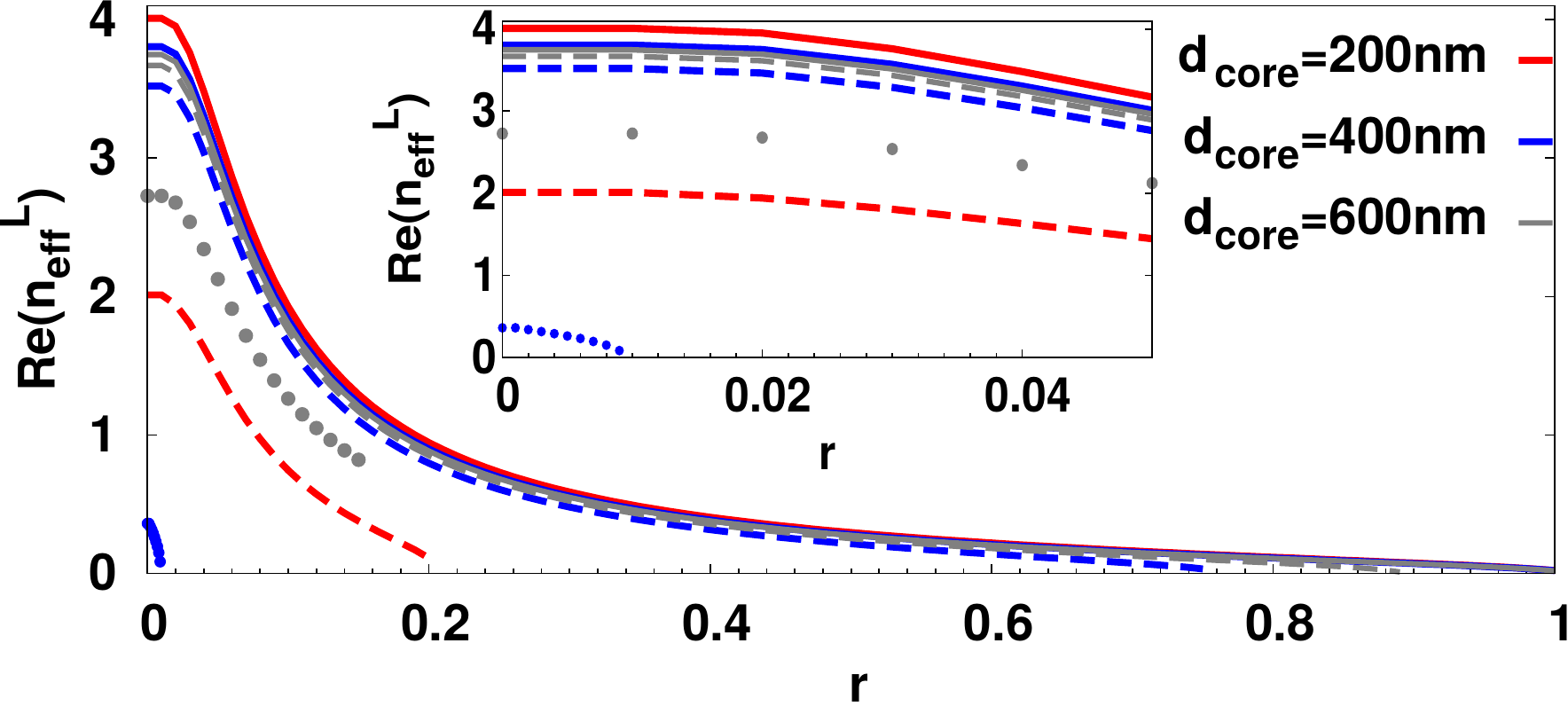}}
\caption{Linear dispersion curves for symmetric NPSWs  as a function of $r$ parameter in the elliptical case for three different core thicknesses $d_{core}$ with $\varepsilon_2= 1.0 \, 10^{-5} + i 0.62$. 
Solid lines stand for  first symmetric modes, dashed lines for first antisymmetric modes, and points for first higher-order symmetric modes. Inset: zoom for the region near $r=0$.} 
\label{fig:linear-dispersion-curve-elliptical}
\end{figure}

To model these anisotropic waveguides,  we assume that the nonlinear Kerr term is isotropic. This is an approximation compared to the full treatment~\cite{Boyd07,elsawy17josab}. To tackle the full case is beyond the scope of our study which is mostly dedicated to the impact of the anisotropy of the linear terms even if its extension can be seen as the next required step in this research field.
In this study, the wavelength is $1.55\, \mu$m. $\varepsilon_{1}$ (corresponding to  a-Si:H), and metal (gold) permittivity are the same as in~\cite{Elsawy16a}, while  $d_{core}$ is fixed to $ 400$nm (except in Fig.~\ref{fig:linear-dispersion-curve-elliptical}), and the  nonlinear coefficient for the first material denoted by $ n_{2,1}$ is set to $ 2. 10^{-17} m^2/W$. 
Next, the two used models of the Kerr nonlinear field dependency  are described.  %

In the first one,  only the transverse component of the electric field $E_x$ which is usually larger than the longitudinal one is taken into account. 
 This approximation has already been used in several models of isotropic NPSWs~\cite{Walasik14a,Walasik15a,Elsawy16a}. It gives similar results than  more accurate approaches where all the electric field components are considered in the optical Kerr effect~\cite{Walasik14a,Walasik15a}.  This first model allows us to use our new semi-analytical approach called EJEM (for Extended Jacobi Elliptical Model which is an extension to the anisotropic case~\cite{elsawy17josab} of our already developed JEM valid for isotropic configurations~\cite{Walasik15a}). This approach will provide insights into the dependency of the effective nonlinearity on the opto-geometrical parameters.
This approximation for the nonlinear term also allows us to use the simple fixed power algorithm in the finite element method (FEM) to compute the nonlinear stationary solutions and their nonlinear dispersion curves~\cite{Rahman90,ferrando03spatial-soliton-pcf,Drouart08,Walasik14} in order to validate our EJEM results.

In the second model, all the electric field components are considered in the nonlinear term, and we need to use the more general FEM approach we developed~\cite{elsawy17josab} to generalize the one-component fixed power algorithm~\cite{Walasik14a} in such structures.

Now, we investigate the elliptical case for the metamaterial nonlinear core NPSWs. We choose for material 2 in the core an ENZ-like one such that $\varepsilon_2= 1.0 \, 10^{-5} + i 0.62$ being similar to the one provided in~\cite{Capretti:15}.
We start this study by the linear case in which the main linear modes we found are of plasmonic type.
 For the metamaterial core, besides the permittivities, we have the ratio $r$ defined above as new degree of freedom. As a result, one can obtain linear dispersion curves as a function of $r$. Fig.~\ref{fig:linear-dispersion-curve-elliptical} shows such curves for several values of the core thickness $d_{core}$.  $n_{eff}^L$ indicates $n_{eff}$ of the linear case.
One can see that it is possible to choose configurations where only the first symmetric mode is kept. This kind of behaviour can be an advantage to reach a simpler and better control of nonlinear propagating solutions as a function of power~\cite{Akhmediev97} or to tune the linear dispersion properties as a function of wavelength to manage the dispersion coefficients.
\begin{figure}[htbp]
\centerline{\includegraphics[width = 1.0\columnwidth,angle=0,clip=true,trim= 0 0 0 0]{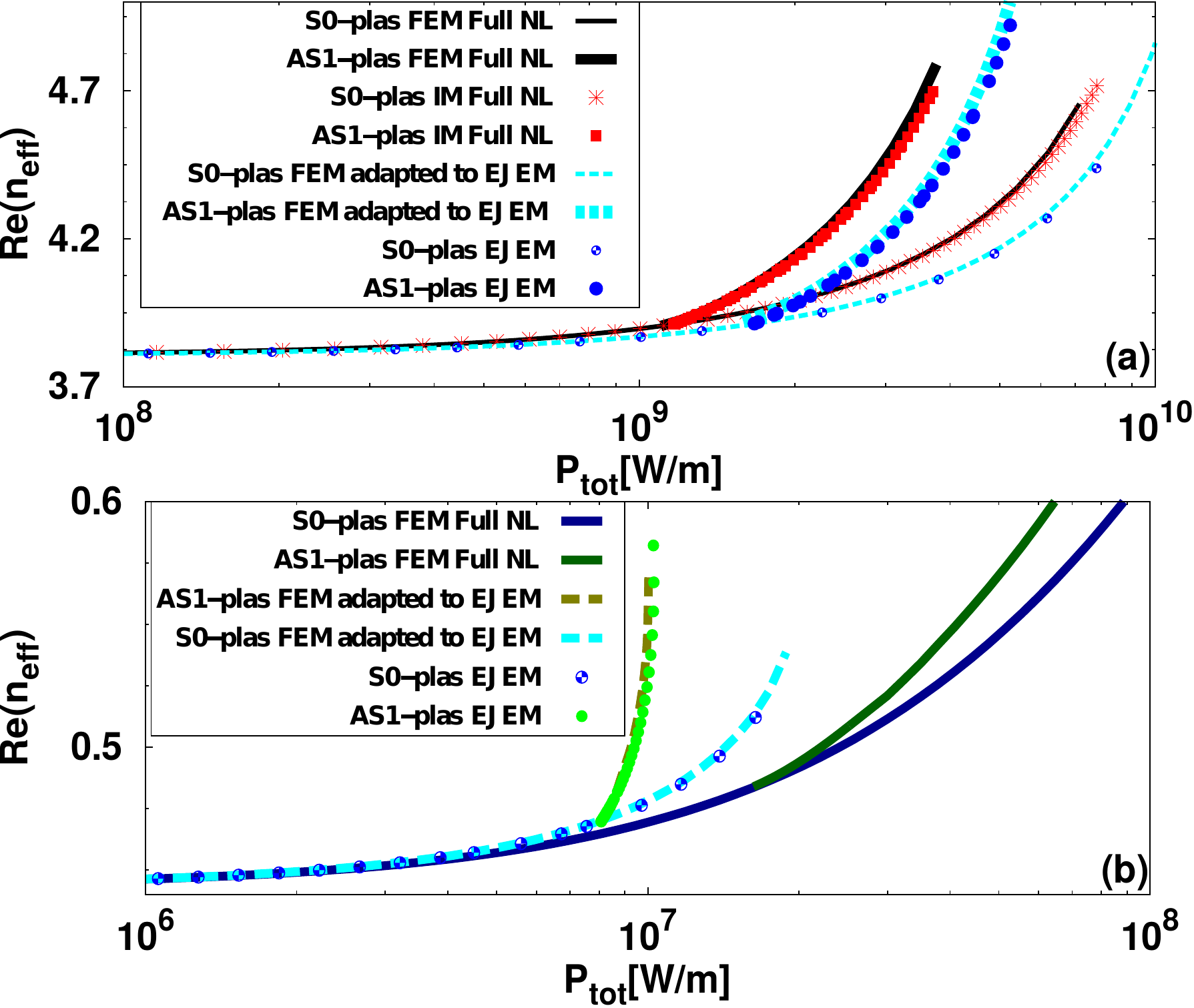}}
\caption{Nonlinear dispersion curves for symmetric NPSWs as a function of total power  $P_{tot}$. Both the symmetric modes (bottom branch for each color) denoted  S0-plas and the asymmetric ones AS1-plas (upper branch after bifurcation) are shown,  the mode notation is fully coherent with the ones used for the simple~\cite{Walasik15a,Walasik15b} or improved~\cite{Elsawy16a} isotropic NPSWs.  
 (a) Isotropic case with the  EJEM, the FEM with and without all the electric field components in the nonlinear term, and the IM.  (b) Elliptical anisotropic case with the EJEM, and the two FEMs.} 
\label{fig:nonlinear-dispersion-curve-elliptical-validation}
\end{figure}

As a test signature for strong nonlinear spatial behaviour and a demanding validity check, we depict the Hopf bifurcation of symmetric mode toward an asymmetric mode in symmetric isotropic and anisotropic NPSWs. In Fig.~\ref{fig:nonlinear-dispersion-curve-elliptical-validation}, we provide the results obtained with the methods we used, the EJEM one and the two FEM ones with and without all the electric field components in the nonlinear term. For comparison with this last case, we also use the interface model (IM) we developed previously to study the isotropic case taking into account all the electric field components~\cite{Walasik15a}.
\begin{figure}[htbp]
\centerline{\includegraphics[width = 1\columnwidth,angle=0,clip=true,trim= 0 0 0 0]{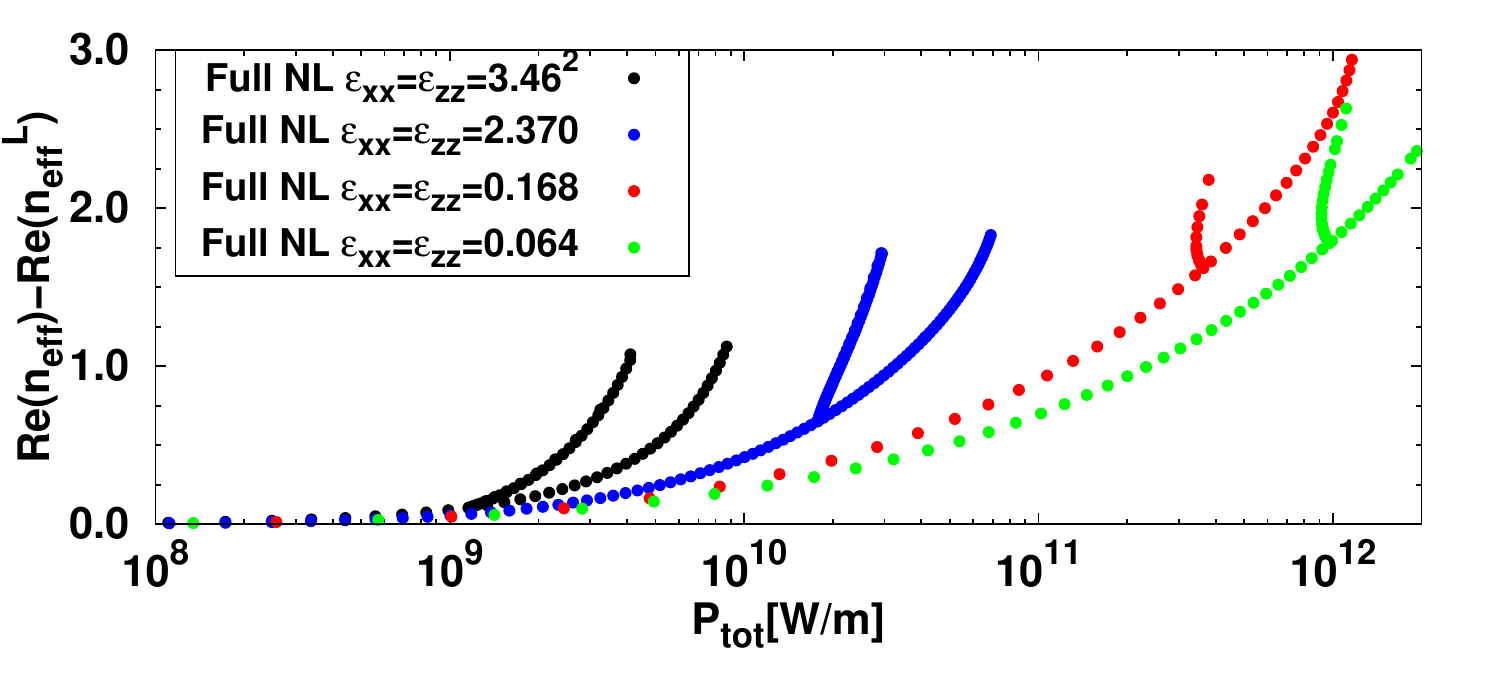}}
\caption{Nonlinear dispersion curves for isotropic symmetric NPSWs as a function of total power  $P_{tot}$  and for different values of linear core permittivity. Both the symmetric modes and the asymmetric ones are shown.}
\label{fig:nonlinear-dispersion-curve-elliptical-isotrope}
\end{figure}
 First for the isotropic case (Fig.~\ref{fig:nonlinear-dispersion-curve-elliptical-validation} (a)),  the FEM taking account only the electric field transverse component is able to recover the results from the EJEM, and our FEM with both electric field components reproduces the results obtained from the IM. Second, for the anisotropic case (Fig.~\ref{fig:nonlinear-dispersion-curve-elliptical-validation} (b)), the EJEM and FEM agrees well. As expected, the results between FEM with and without all the electric field components in the nonlinear term differ slightly at high powers. Consequently, these results prove the validity of our numerical methods for nonlinear studies including the anisotropic case.

Despite,  the enhancement of nonlinear effects due to the use of ENZ materials demonstrated both theoretically~\cite{ciattoni_extreme_2010,neira_eliminating_2015,ciattoni_enhanced_2016}   and experimentally~\cite{Alam795}, Fig.~\ref{fig:nonlinear-dispersion-curve-elliptical-isotrope} shows that, in the isotropic case, the ENZ material core does not reduce the bifurcation threshold  but increases it. This can be understood qualitatively as follows.
In ENZ material the wavelength light is stretched thus the two core interfaces are then more tightly coupled and more power is needed by the nonlinearity to break the symmetry of the field profile. 

\begin{figure}[htbp]
\centerline{\includegraphics[width = 1\columnwidth,angle=0,clip=true,trim= 0 0 0 0]{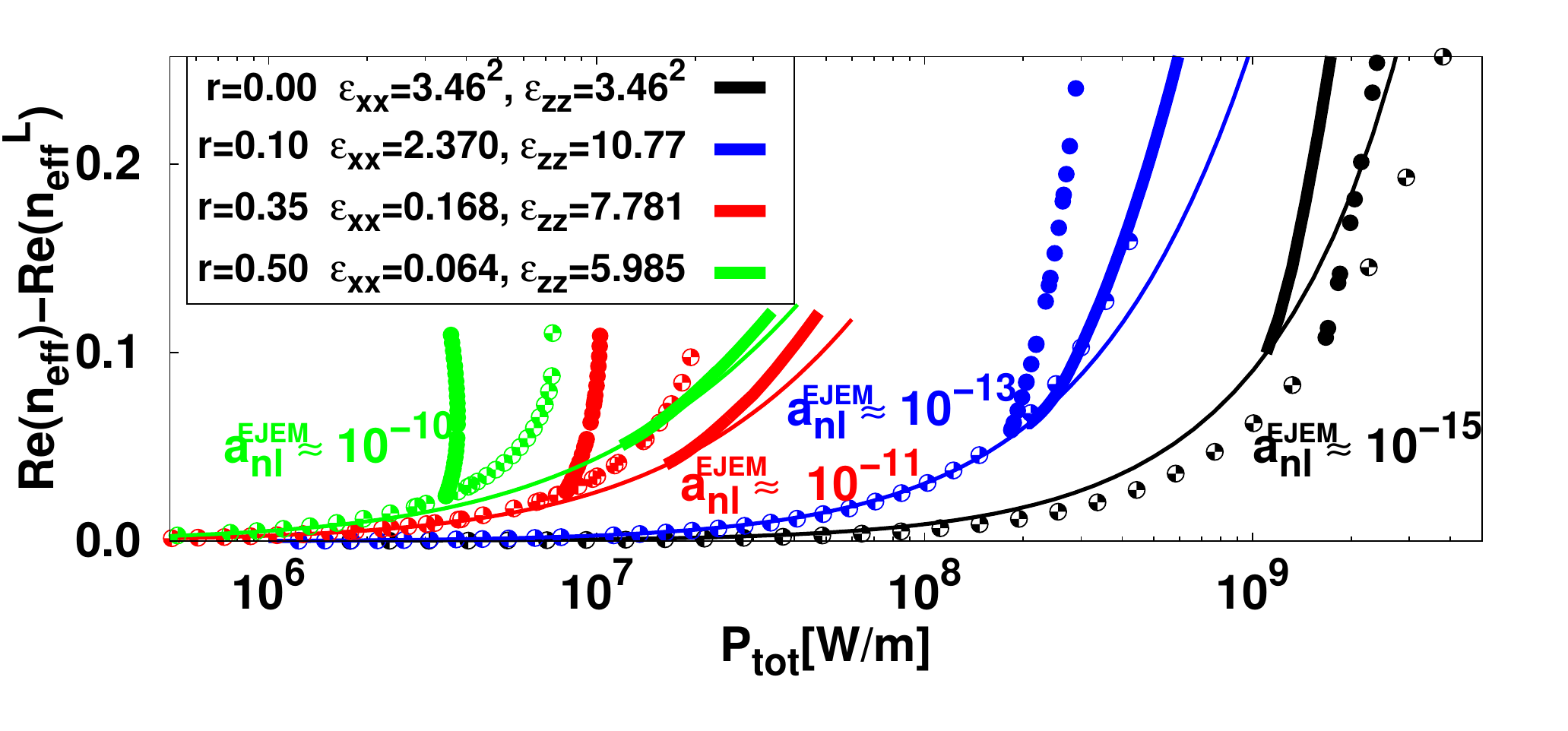}} 
\caption{Nonlinear dispersion curves for elliptical anisotropic and isotropic symmetric NPSWs as a function of total power  $P_{tot}$ for different values of the ratio $r$. Both the symmetric modes and the asymmetric ones are shown. The  curves have been translated along the $y$-axis to improve visibility. The associated  values of the effective nonlinearity $ a_{nl}^{\textsc{ejem}}$ at the bifucation threshold are also given.} 
\label{fig:nonlinear-dispersion-curve-elliptical-anisotrope}
\end{figure}

\begin{figure}[htbp]
\centerline{\includegraphics[width = 1\columnwidth,angle=0,clip=true,trim= 0 0 0 0]{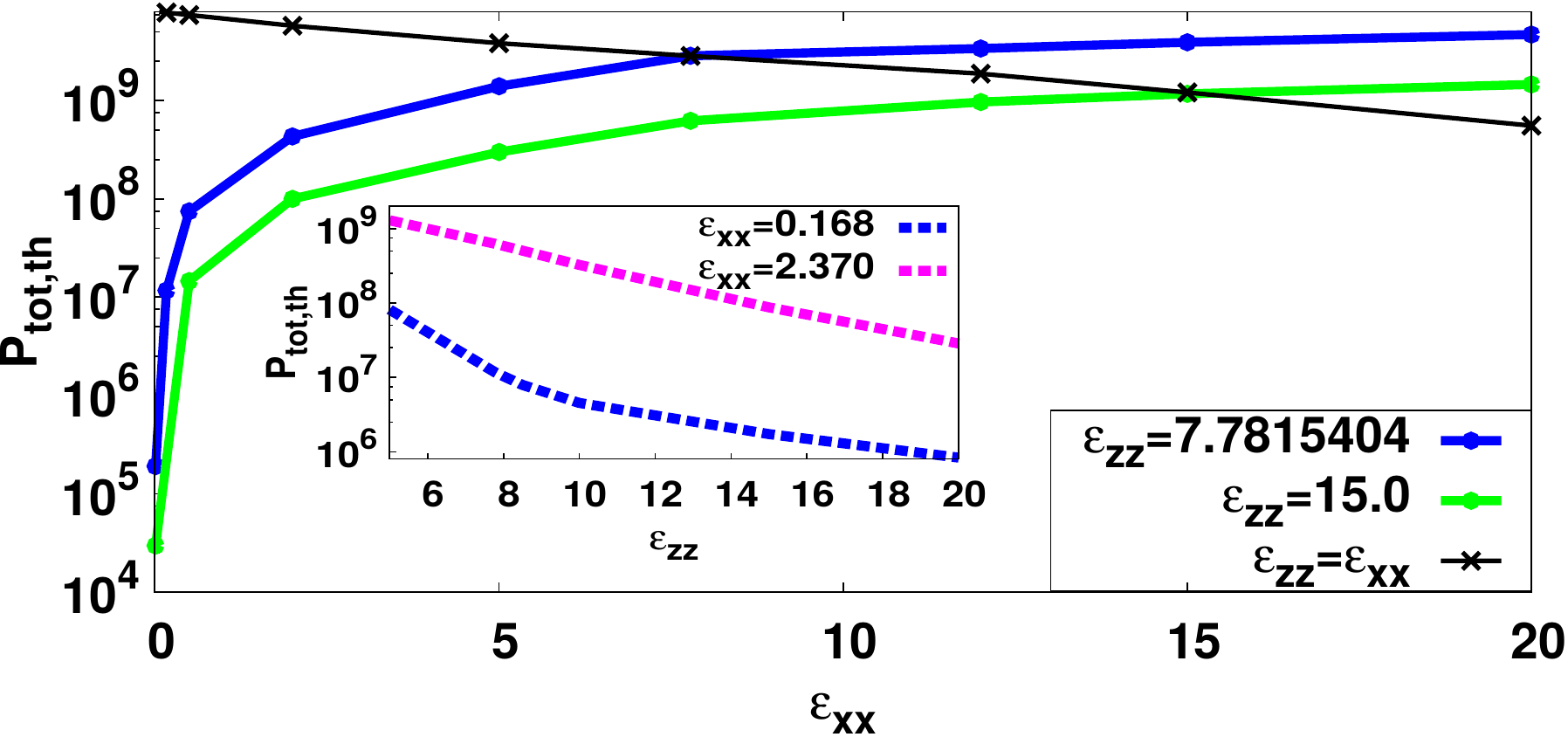}} 
\caption{Power threshold $P_{tot,th}$ as a function of linear transverse permittivity  $\varepsilon_{xx}$  in the elliptical case  for two longitudinal permittivity   $\varepsilon_{zz}$ values.  Isotropic case is shown by the black curve.  Inset:  $P_{tot,th}$ as a function of $\varepsilon_{zz}$ for two  $\varepsilon_{xx}$ values.}
\label{fig:elliptical-case-anisotry-quantitative-effects}
\end{figure}

In the anisotropic case, as it can be seen in Figs.~\ref{fig:nonlinear-dispersion-curve-elliptical-validation} (b) and ~\ref{fig:nonlinear-dispersion-curve-elliptical-anisotrope},  to consider a nonlinear core with ENZ $\varepsilon_{xx}$ and large $\varepsilon_{zz}$  allows us to drastically reduce  the needed total power to induce the symmetry breaking in the NPSW compared to the usual isotropic case.  As a result, we can shift from a GW/m threshold to approximatively 50 MW/m one. 
Using our semi-analytical EJEM, we obtain the following analytical expression for the effective nonlinearity term~\cite{Walasik15a} in the studied anisotropic NPSWs~\cite{elsawy17josab}:
\begin{equation}
  \label{eq:effective-nonlinearity}
  a_{nl}^{\textsc{ejem}} =  -\tilde{\alpha} n_{eff}^2   \left( n_{eff}^2  \left(\varepsilon_{xx}- \varepsilon_{zz}\right) - \varepsilon_{xx}^2 \right) /  \left( \varepsilon_{xx}^4 c^2 \varepsilon_0^2 \right)
\end{equation} 
with $ \tilde{\alpha} = \varepsilon_0 c \Re e(\varepsilon_1) (1- r) n_{2,1}$. Consequently, for the NPSWs, the reinforcement of the effective nonlinearity when ENZ $\varepsilon_{xx}$ and large $\varepsilon_{zz}$ is clearly understood and quantified.  It seems to have been partially overlooked in some previous studies due to the fact that most attention was dedicated to the permittivity tensor case one  $[ \varepsilon_{//}  \;\,  \varepsilon_{\perp}   \;\,  \varepsilon_{//}] $ leading to $\varepsilon_{xx}=\varepsilon_{zz}=\varepsilon_{//}$, and not the case two as studied here with $[ \varepsilon_{\perp}  \;\,  \varepsilon_{//} \;\, \varepsilon_{//}] $ leading to non-vanishing terms in \eqref{eq:effective-nonlinearity}.
The observed reduction of the bifurcation threshold is not possible neither in isotropic improved NPSWs~\cite{Elsawy16a} nor in isotropic ENZ NPSWs as  shown in Fig.~\ref{fig:nonlinear-dispersion-curve-elliptical-isotrope}. Fig.~\ref{fig:elliptical-case-anisotry-quantitative-effects}  gives the thresholds as a function of transverse and longitudinal permittivities for several configurations. Above the black line associated to the isotropic case, one gets higher thresholds while they are smaller below. In the anisotropic case, for ENZ  $\varepsilon_{xx}$, one can  see the strong decrease of the threshold. For a fixed  $\varepsilon_{xx}$, an increase of  $\varepsilon_{zz}$ induces a decrease of the bifurcation threshold (see inset in Fig.~\ref{fig:elliptical-case-anisotry-quantitative-effects}).
Nevertheless, it can be argued that order of magnitude threshold decreases have already been predicted~\cite{Walasik15a,Walasik15b} but this result was obtained using a large increase of the core size moving the structure from nanophotonics to large integrated optics structures. In the present case, small core thicknesses can be kept allowing not only a limited footprint for the devices but also a limited number of propagating modes in the metamaterial based NPSWs, eventually only the fundamental symmetric mode (see Fig.~\ref{fig:linear-dispersion-curve-elliptical}) and the associated  asymmetric one. One can also notice that the slopes of the symmetric mode nonlinear dispersion curves for the studied highly anisotropic NPSWs are not negligible even below the reduced bifurcation threshold involving important nonlinear effects on the propagation of this mode even at lower powers. Another consequence of the use of a highly anisotropic elliptical metamaterial core is the low value of the effective indices for the main modes (see Fig.~\ref{fig:nonlinear-dispersion-curve-elliptical-validation} (b)) ensuring a slow light enhancement for the nonlinear effects in temporal propagation configurations~\cite{baba08slow-light-PC}. %
The impact of the core anisotropy is also seen on the dispersion curve of the main asymmetric mode. As shown, in Fig.~\ref{fig:nonlinear-dispersion-curve-elliptical-isotrope} (isotropic case), the lower the index core permittivity, the larger the slope of the  asymmetric mode branch. Moreover, for ENZ isotropic core ( $\varepsilon_{core} \lesssim 1$),  the slope is negative   
while, as shown in Fig.~\ref{fig:nonlinear-dispersion-curve-elliptical-anisotrope}, for highly anisotropic core with ENZ $\varepsilon_{xx}$ and large $\varepsilon_{zz}$, the slope of the asymmetric mode near the bifurcation point stays positive. 
 If we assume that the stability results we obtained for isotropic NPSWs~\cite{Walasik15b} can be extended to the anisotropic case, these two features suggest that the asymmetric mode should be instable in isotropic ENZ core NPSWs ( $\varepsilon_{core} \lesssim 1$) while the same mode should be stable  for highly anisotropic core with ENZ $\varepsilon_{xx}$ and large $\varepsilon_{zz}$ (a full stability study of the main modes as described in~\cite{Walasik15b} for simple NPSWs is beyond the scope of this work).  

\begin{figure}[htbp]
\centerline{\includegraphics[width = 1.0\columnwidth,angle=0,clip=true,trim= 0 0 0 0]{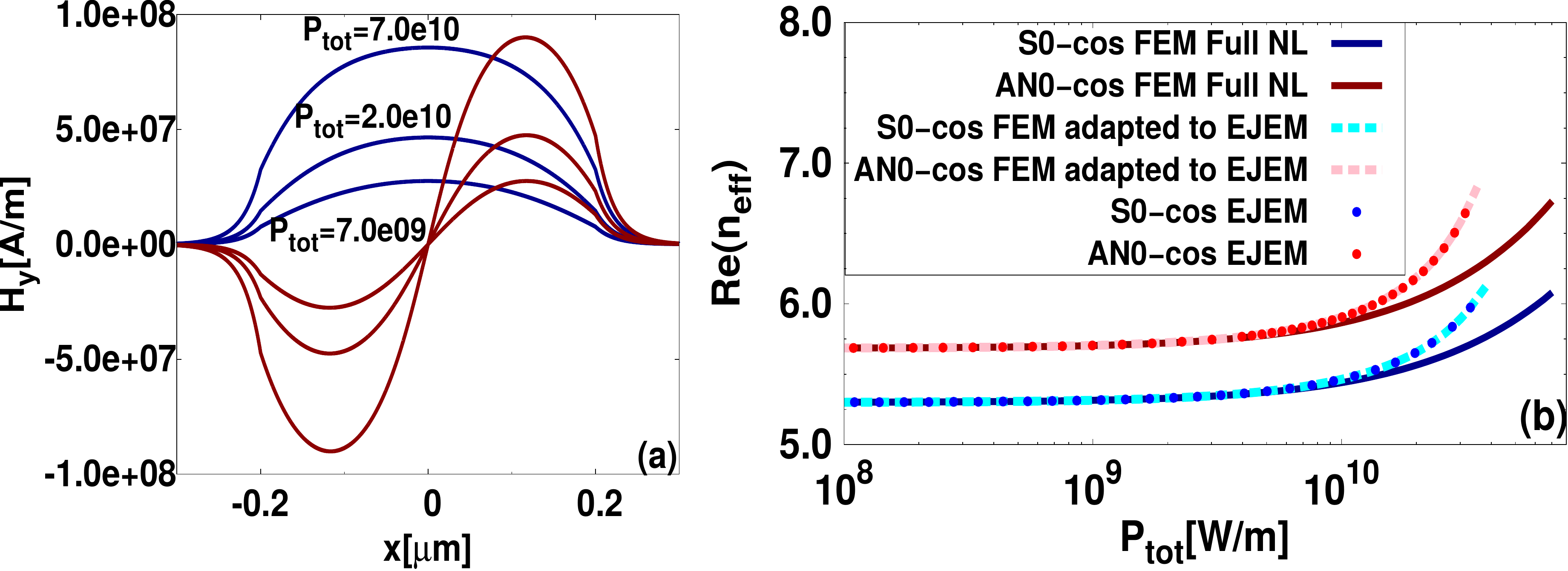}}

\caption{Hyperbolic symmetric NPSWs  for $\varepsilon_{xx}=26.716$ and $ \varepsilon_{zz}=-51.514$ (obtained from material 2  permittivity  $\varepsilon_{2} = -115 + i6 $ (copper) and $r=0.5$). (a) Field profile $H_y(x)$  as a function of total power $P_{tot}$. (b) Nonlinear dispersion curves  as a function of  $P_{tot}$.  The main symmetric and  antisymmetric  modes are shown for both the EJEM and the two FEM.} 
\label{fig:field-profiles-hyperbolic-case}     
\end{figure}

We now investigate the hyperbolic case where the metamaterial core is such that $\varepsilon_{xx} >0 $ and  $\varepsilon_{zz}<0$. In this case, it is known that non-local effects  can be neglected in EMT  as soon as the condition $d_1=d_2$ is fulfilled corresponding to $r=0.5$~\cite{chebykin_nonlocal_2011}. We will limit the study to such configurations.  
Linear studies of waveguides involving such linear metamaterial core have already been published~\cite{avrutsky_highly_2007}. For NPSWs, we found that the main modes are core localized unlike the ones of simple NPSWs, and that the effective nonlinearity can be negative for the investigated modes meaning  that the initial positive Kerr nonlinearity can finally act as a negative one in such anisotropic configuration. This can be understood looking at Eq.~(\ref{eq:effective-nonlinearity}). Fig.~\ref{fig:field-profiles-hyperbolic-case} (a)  illustrates this phenomenon. We also found that the asymmetric mode we can obtain as a mathematical solution of the nonlinear dispersion equation is actually unbounded~\cite{elsawy17josab}, knowing that similar unbounded modes were already obtained in other nonlinear structures~\cite{Chen88}. Therefore, this asymmetric mode can not be considered as an acceptable solution of our physical problem. %
The nonlinear dispersion curves of the main symmetric and antisymmetric modes are given in Fig.~\ref{fig:field-profiles-hyperbolic-case} (b).
Once again, one can see the crucial influence of the metamaterial core properties on the type and behaviour of the propagating nonlinear solutions.

 The found spatial nonlinear effects are a signature of a strong nonlinear reinforcement.  We move from a GW/m bifurcation threshold required in the isotropic cases~\cite{Walasik15b} even in improved NPSWs~\cite{Elsawy16a} to  tens of MW/m one for elliptical anisotropic NPSWs with ENZ $\varepsilon_{xx}$ and large $\varepsilon_{zz}$. This improvement makes the properties of the proposed waveguides really achievable to  materials used in current fabrication processes in photonics and also to most characterization setups. 

\noindent \textbf{\large Acknowledgments.} G.~R. would like to thank the PhD school ED 352 "Physique et Sciences de Matière" and the International Relation Service of Aix-Marseille University (project AAP 2015 "Noliplasmo 2D")  for their respective fundings.

\clearpage

\end{document}